\documentclass[12pt, amssymb, nofootinbib, aps, preprintnumbers, prd]{revtex4}

\usepackage[english]{babel}
\usepackage{amsmath}
\allowdisplaybreaks
\usepackage{amssymb}
\usepackage{amsbsy}
\usepackage{amstext}
\usepackage{graphicx}
\usepackage{float}
\makeatletter\AtBeginDocument{\let\@elt\relax}\makeatother
\usepackage{hyperref}
\usepackage{pst-node}
\usepackage{verbatim}
\usepackage{tikz}
\usepackage{cancel}
\usepackage{braket}
\usepackage[normalem]{ulem}
\usetikzlibrary{snakes,decorations.pathmorphing}

\newcommand{\be}{\begin{eqnarray}}
\newcommand{\ee}{\end{eqnarray}}
\newcommand{\bdm}{\begin{displaymath}}
\newcommand{\edm}{\end{displaymath}}
\newcommand{\ds}{\displaystyle}

\newcommand{\nn}{\nonumber}
\newcommand{\ba}{\begin{array}}
\newcommand{\ea}{\end{array}}
\newcommand{\pa}[1]{\left(#1\right)}
\newcommand{\paq}[1]{\left[#1\right]}

\newcommand{\K}{{\bf k}}
\newcommand{\Q}{{\bf q}}

\newcommand{\X}{{\bf x}}

\newcommand{\II}{{\rm{I}}}
\newcommand{\JJ}{{\rm{J}}}

\begin{document}

\title{Gravitational memory contributions to waveform and effective action}

\author{Gabriel Luz Almeida}
\email{galmeida@ustc.edu.cn}
\affiliation{Interdisciplinary Center for Theoretical Study, University of Science and Technology of China\\ and Peng Huanwu Center for Fundamental Theory,\\
Hefei, Anhui 230026, China}

\author{Alan M\"{u}ller}
\email{alan.muller@unesp.br}
\affiliation{Instituto de F\'\i sica Te\'orica, UNESP - Universidade Estadual Paulista, Sao Paulo 01140-070, SP, Brazil}

\author{Stefano Foffa}
\email{stefano.foffa@unige.ch}
\affiliation{D\'epartement de Physique Th\'eorique and Gravitational Wave Science Center, Universit\'e de Gen\`eve, CH-1211 Geneva, Switzerland}

\author{Riccardo Sturani}
\email{riccardo.sturani@unesp.br}
\affiliation{Instituto de F\'\i sica Te\'orica, UNESP - Universidade Estadual Paulista, Sao Paulo 01140-070, SP, Brazil \\ ICTP South American Institute for Fundamental Research, Sao Paulo 01140-070, SP, Brazil}

\begin{abstract}
  We use Effective Field Theory techniques to derive the quadrupole-quadrupole part of the gravitational wave, obtaining a waveform in agreement with
    previous results found within the multipolar-post-Minkowskian method.
      This result is obtained by carefully avoiding implicit coordinate transformation, that would also alter the form of the radiation reaction force in the harmonic gauge.
An in-in effective action is then derived along the same principles and it is
shown to provide energy and angular momentum balance equations  consistent with the corresponding fluxes carried at infinity by
gravitational radiation.

\end{abstract}


\preprint{USTC-ICTS/PCFT-24-34}

\maketitle


\section{Introduction}
Gravitational wave production from an isolated source, and the subsequent
backreaction on the source itself, has been a major subject of interest since the early times of the investigation of the two body problem in General Relativity.

If at the lowest, linear level, the subject is well understood and systematized \cite{Thorne:1980ru,Thorne:1969rba,Burke:1970wx}, and presents similarities with the analog problem in electrodynamics,
things become more complicated and interesting when the non-linear nature of gravity is taken into account.

Tails \cite{Blanchet:1987wq}  and more in general hereditary effects \cite{Christodoulou:1991cr,Blanchet:1992br}, that is the interaction of gravitational radiation with other components of the gravitational field generated by the same source
(such as the quasi-static curvature for the tails, or another gravitational wave in the case of the so-called memory interactions), has attracted scholars' attention since almost four decades, and do not cease to surprise us, especially for their impact on the two-body dynamics. 

Compact binary mergers are the prominent target of present \cite{TheLIGOScientific:2014jea, TheVirgo:2014hva} and future gravitational waves detectors \cite{Maggiore:2019uih,Reitze:2019iox,LISA:2022kgy},
and high-precision analytical determination of the waveform in the inspiral phase is one of the main tools to build templates for matched filtering detection. The gravitational wave phase is currently known at fourth and an half post-Newtonian level of precision ($4.5$PN) \cite{Blanchet:2023bwj}, that is up to ${\cal O}\pa{v/c}^9$, or equivalently ${\cal O}\pa{Gm/r}^{9/2}$\footnote{$r$ and $v$ are,
respectively, the relative distance and velocity of the binary constituents
and $m$ the total mass.},
for the case of spin-less binary components, and more accurate templates
  are required for future detectors \cite{Purrer:2019jcp} whose loudest signals are expected to
  have signal-to-noise ratio 10-100 times higher than the present loudest event.

Tail interactions, non-local in time, play a major role already at $1.5$PN in the emitted radiation, and at $4$PN in the binary dynamics \cite{Foffa:2011np,Damour:2014jta,Galley:2015kus} where for the first time the impact of emitted radiation becomes manifest even in the conservative sector.
Memory terms appear at 2.5PN in the waveform (under the form of quadrupole-quadrupole radiation \cite{Blanchet:1997ji}) and at $5$PN, that is the next precision step beyond the present state-of-the-art, in the binary dynamics. Despite giving a purely local-in-time contribution to the radiated energy flux, memory effects shall not be viewed as easier-to-handle than non-local tails: for instance, conservative and dissipative effects are mixed together in such a way to require a systematic use of the in-in \cite{Galley:2009px} formalism, which could be avoided (with some precautions) in the study of tails. 

In addition to that, the first attempts to deal with such effects \cite{Foffa:2019eeb,Foffa:2021pkg}, and to combine them with quasi-static (``potential") and tail contributions into a comprehensive $5$PN effective action
 \cite{Foffa:2019hrb,Blumlein:2019zku,Blumlein:2020pyo,Foffa:2020nqe,Almeida:2021xwn, Almeida:2023yia} fell short to provide the expected scattering behaviour, as it was first invoked from scaling arguments \cite{Bini:2019nra,Bini:2021gat} and then actually computed \cite{Dlapa:2022lmu} in a post-Minkowskian setting (that is a perturbative expansion in Newton's constant only, as appropriate in a scattering problem).
 The latter point has been recently addressed in \cite{Porto:2024cwd} where, among other things, a self-consistent procedure had been introduced to compute (i) the quadrupole-quadrupole emission, (ii) the corresponding effective action and (iii) a memory contribution to the scattering angle which combines with other contributions to give correct value.
 
It is worth noting that the waveform as well as the energy and angular momentum fluxes as derived in \cite{Porto:2024cwd} appear in a different form with respect to the long-established results obtained from the multipolar post-Minkowskian (MPM) formalism \cite{Blanchet:1996wx,Blanchet:1997ji,Arun:2007sg}. 
This is reminiscent of a similar discrepancy between the waveform derived from modern amplitude methods and the result of the MPM derivation, which was eventually resolved by recognizing in the former a contribution equivalent to a frame rotation \cite{Bini:2024rsy,Georgoudis:2024pdz}.

The present work brings clarification about this point: by using an EFT-inspired approach \cite{Goldberger:2004jt} similar to \cite{Porto:2024cwd} (but with some crucial differences), we compute from first principles a quadrupole-quadrupole waveform which matches identically the MPM result. We understand the relation between the latter and the result reported in \cite{Porto:2024cwd} as due to a near-zone coordinate transformation, which induces a shift in the relative coordinates. Such shift is the same that relates
  two forms of the radiation reaction force: Burke-Thorne, usually employed in the EFT treatment, and Damour-Deruelle \cite{Damour:1981bh}, derived in the
  \emph{harmonic gauge}, as defined in \cite{Blanchet:2000nv}, and usually employed in the MPM formalism.\footnote{In \cite{Arun:2009mc} the original harmonic coordinates have been relabeled \emph{standard}, to distinguish them from their \emph{modified} version \cite{Arun:2007rg}, which get rid of gauge-dependent logarithms in the 3PN equation of motions. The distinction between standard and modified harmonic coordinates is irrelevant for the present paper.}
  
The plan of the paper is the following: in section \ref{sec:setup} we describe our framework and derive leading order results, including the radiation reaction force in the Damour-Deruelle form. Then in section \ref{sec:WF} we proceed to derive the quadrupole-quadrupole waveform, we discuss the (not straightforward) comparison with the MPM literature and we clarify the relation with the form appearing in the recent EFT result.
The following section \ref{sec:Sinin} contains our derivation of the effective action and the proof of consistency with the waveform, and we draw our conclusions in the final section \ref{sec:disc}. In Appendix \ref{app:magLtail}, we provide a self-consistent picture on an issue associated to emission process of the angular momentum failed tail investigated in our previous paper \cite{Almeida:2023yia}, in light of the approach discussed in the present work. Appendic \ref{app:Schott} contains the expressio. of Schott terms in the standard harmonic gauge.

\section{Setup and leading order results}
\label{sec:setup}

As a start, we summarize here some notions used in our work, building from the fundamental ones
up to some nontrivial considerations on radiation reaction. 

\subsection{Notation}
$T_{\mu\nu}$ is the energy momentum tensor of a binary system,\\
$M$ is the 0-moment of  energy of the binary,\\
$M_{i}$ is the first moment of $T_{00}$ quadrupole moment,\\
$M_{ij}$ is the second moment of $T_{00}$, i.e. the quadrupole moment, with trace,\\
$Q_{ij}$ is the traceless quadrupole moment,\\
$P^i$ is the 0-moment of $T_{0i}$ and $P^{i,j}$ the first moment,\\
${\bf r}$ is the vector distance between binary consistuents, 
$R$ is the source-observer distance.

\subsection{Waveform as one point function}
The dynamics of the gravitational field is determined by the Einstein-Hilbert action plus a gauge-fixing term:
\begin{align}\label{bulkgravity}
    {\cal S}_{\rm EH}^{\rm GF} &= \frac{1}{16\pi G_d} \int {\rm d}^{d+1}x \sqrt{-g} \left[ R(g) - \frac12 \Gamma_\mu \Gamma^\mu \right]\,.
\end{align}

If we denote by $i\mathcal{A}_{\alpha\beta}$ the amplitude for the emission of the gravitational perturbation
$h^{\alpha\beta}$, the classical field at a spacetime position $x$ far from the source,
given by the one-point function $\braket{h_{\mu\nu}(x)}$, can be written as\footnote{We use the notation $\int_y=\int {\rm d} y$, if $y$ is a direct space variable,
and $\int_q=\int \frac{{\rm }d q}{2\pi}$ if $q$ is a Fourier space one.}
\begin{equation}
\label{eq:hexpect}
\braket{h_{\mu\nu}(x)} = \int\mathcal{D}h\, e^{iS[h]}h_{\mu\nu}(x)=32 \pi G_d \int_\K \frac{{\rm d}\omega}{2\pi} \frac{e^{-i\omega t+i\K\cdot\X}}{\K^2-(\omega+i\epsilon)^2}
{{\cal P}_{\mu\nu}}^{\alpha\beta}\tilde{\cal A}_{\alpha\beta}(\omega,\K)\,,
\end{equation}
$G_d$ being $d+1$-dimensional gravitational constant (which reduces to Newton
constant $G$ for $d=3$),
${{\cal P}_{\mu\nu}}^{\alpha\beta}\equiv\frac12\pa{\delta_\mu^\alpha\delta_\nu^\beta+\delta_\mu^\beta\delta_\nu^\alpha-\frac{2}{d-1}\eta_{\mu\nu}\eta^{\alpha\beta}}$, and
the appropriate, retarded boundary condition has been inserted.

Expressing $\tilde{\cal A}_{\alpha\beta}$ in direct space and integrating
over $\omega,\K$, eq.~(\ref{eq:hexpect}) becomes
\be
\label{eq:hradfar}
\braket{h_{\mu\nu}(x)}=8 G \int_{\X'}\frac{{{\cal P}_{\mu\nu}}^{\alpha\beta}{\cal A}_{\alpha\beta}\pa{t-|\X-\X'|,\X'}}{|\X-\X'|}\rightarrow\ds\frac{8 G}{R} \int_\omega {{\cal P}_{\mu\nu}}^{\alpha\beta}{\cal A}_{\alpha\beta}\pa{\omega,\omega {\bf n}}{\rm e}^{-i\omega\pa{ t-R}}\,,
\ee
where in the last passage $|\X-\X'|\to R$ in the denominator
  and $|\X-\X'|\to R-{\bf{n}}\cdot\X'$ in the ${\cal A}_{\alpha\beta}$
  time argument, as per standard textbook procedure.

The linearized Lorentz gauge condition translates into ``Ward'' identities
for the classical process consisting of the emission of a single gravitational
mode\footnote{Note that the trace reversion operator ${{\hat{\cal P}}_{\mu\nu}}^{\alpha\beta}\equiv \frac 12 \pa{\delta_\mu^\alpha\delta_\nu^\beta+\delta_\mu^\beta\delta_\nu^\alpha-\eta_{\mu\nu}\eta^{\alpha\beta}}$, which turns $h_{\alpha\beta}$ into $\bar{h}_{\mu\nu}$, is identical to its inverse ${{\cal P}_{\mu\nu}}^{\alpha\beta}$ only for $d=3$.}
\begin{equation}\label{eq:Lorentz}
  \partial^\mu\braket{\bar{h}_{\mu\nu}(x)} = 0 \quad \Leftrightarrow \quad
  i k^\mu{\cal A}_{\mu\nu}(\omega,\K) = 0\quad \pa{k^\mu\equiv\{\omega,k^i\}}\,.
\end{equation}

\subsection{Linear coupling and matching}

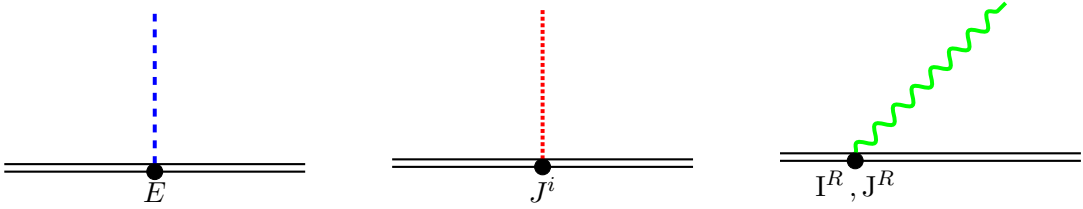
\begin{figure}
	\begin{center}
		\begin{tikzpicture}
      			\draw [black, thick] (0,0) -- (4.,0);
      			\draw [black, thick] (0,-0.1) -- (4.,-0.1);
      			\draw [blue, dashed, line width=1.5] (2,0) -- (2,2);
      			\filldraw[black] (2.,-0.1) circle (3pt) node[anchor=north] {$E$};
		\end{tikzpicture}\hspace{1cm}
		\begin{tikzpicture}
      			\draw [black, thick] (0,0) -- (4.,0);
      			\draw [black, thick] (0,-0.1) -- (4.,-0.1);
      			\draw [red, densely dotted, line width=1.5] (2,0) -- (2,2);
      			\filldraw[black] (2.,-0.1) circle (3pt) node[anchor=north] {$J^i$};
		\end{tikzpicture}\hspace{1cm}
		\begin{tikzpicture}
      			\draw [black, thick] (0,0) -- (4.,0);
      			\draw [black, thick] (0,-0.1) -- (4.,-0.1);
    			\draw[decorate, decoration=snake, line width=1.5pt, green] (1,0) -- (3,2) ;
      			\filldraw[black] (1.,-0.1) circle (3pt) node[anchor=north] {$\II^R\,, \JJ^R$};
		\end{tikzpicture}\hspace{1cm}
	\end{center}
	\caption{Leading order emission amplitudes. Blue dashed line
          (sourced by the total mass $E$ of the system) refers
          to a $h_{00}$ polarization, dotted red to $h_{0i}$ (sourced here
          by angular momentum $J_i$). Wavy green line refers to a radiative
          mode, sourced by generic electric and magnetic multipole, respectively
        denoted by $\II^R,\JJ^R$.}
	\label{fig:simple}
\end{figure}

The linear part of the matter-gravity coupling can be derived from the long-wavelength expansion of the minimal fundamental coupling of the gravitational perturbation to the energy-momentum tensor of the source:
\be
\label{eq:mult_equiv}
{\cal S}_{\rm source} &=&\ds \frac12 \int_{t\,,\X} T^{\mu\nu}h_{\mu\nu} \simeq \frac12\ds \sum_r \frac1{r!}\int_{t\,,\X}  T^{\mu\nu}x^R\partial_R h_{\mu\nu}\,.
\ee
Considering just the first terms of the expansion, this gives
\be\label{eq:moments_action}
{\cal S}_{\rm source}&\simeq&\frac12\int_t \paq{M h_{00}+ M^i h_{00,i}+\frac12 M^{ij}h_{00,ij}+2 P^i h_{0i}+\pa{2P^{(i,j)}-L^{i|j}}h_{0i,j}+S^{ij}h_{ij}}\,,
\ee
with
\be
&&M\equiv\int_\X T^{00}\,,\quad M^i\equiv\int_\X T^{00}x^i\,,\quad M^{ij}\equiv\int_\X T^{00}x^i x^j\,,\\
&& P^i\equiv\int_\X T^{0i}\,,\quad P^{i,j}\equiv\int_\X T^{0i}x^j\,,\quad L^{i|j}\equiv -2\int_\X T^{0[i}x^{j]} \equiv\epsilon_{ijk}L^k\,,\\
&& S^{ij}\equiv  \int_\X T^{ij}\,.
\ee

Keeping only the leading radiative multipole moment,
the corresponding simple emission amplitudes read
\be
\label{eq:LOnoW}
{\cal A}^{\rm LO}_{00}(\omega,\K)&\simeq&\frac M2-\frac i2 k_kM^k-\frac14 k^i k^j M_{ij} \,,\\
{\cal A}^{\rm LO}_{0i}(\omega,\K)&\simeq&-\frac12 P^i+\frac i2 k_kP^{(i,k)}-\frac i4 k_k L^{i|k} \,, \\
\label{eq:AijLO}
{\cal A}^{\rm LO}_{ij}(\omega,\K)&\simeq&\frac12S_{ij}\,.
\ee 
The Ward identities give
\be
\label{eq:WardLO0}
{\rm time\ component:}&&\quad\dot{M}=0\,,\quad \dot{M}^i=P^i\,,\quad \dot{M}^{ij}=2P^{(i,j)}\,,\\
\label{eq:WardLOk}
{\rm spatial\ component:}&&\quad\dot{P}^i=0\,,\quad \dot{P}^{(i,j)}=S^{ij}\,,\quad \dot{L}^{i|j}=0\,;
\ee
 they express the conservation of the energy-momentum tensor of the source, $T^{\mu\nu}_{\ \ ,\nu}=0$, at $O(G^0)$ order.

When replaced back into the LO amplitudes, in the center of mass frame where
$P^i=0=M^i$, they give
\be
\label{eq:LOWard00}
\ds \mathcal{A}^{\rm LO}_{00}  (\omega,\K)&\simeq&\ds \frac12 M(\omega)-\frac14 k^i k^j M^{ij}(\omega)\,,\\
\label{eq:LOWard0k}
\ds \mathcal{A}^{\rm LO}_{0k} (\omega,\K)&\simeq&\ds \frac i4 k_j\epsilon_{ijk}L_i(\omega)+\frac14 \omega k^j M^{jk}(\omega)\,,\\
\label{eq:LOWardkl}
\ds \mathcal{A}^{\rm LO}_{kl} (\omega,\K)&\simeq&\ds -\frac14\omega^2  M^{kl}(\omega) \,,
\ee
The same result can be obtained from the first terms of the following far zone Lagrangian\footnote{$J^{b|iRa}$'s are the $d$-dimensional generalizations \cite{Henry:2021cek} of the 3-dimensional current multipoles $J^{ijR}=\frac12 \epsilon_{ba(i} \left. J^{\underline{b}|jR)a}\right|_{d=3}$, while the $I^{ijR}$'s are the usual completely symmetric mass multipoles. We use $R\equiv i_1\dots i_r$ as a collective index. Indices among round brackets are symmetrized, among square ones antisimmetrized: $A^{(ij)}=\frac 12(A^{i,j}+A^{j,i})$, $A^{[i,j]}=\frac 12(A^{i,j}-A^{j,i})$.}
\be\label{sourcelag}
{\cal S}_{\rm far} &=& \int_t\,\left[ \frac12 E h_{00} -\frac12 J^{b|a} h_{0b,a} - \sum_{r\ge 0}\left( c_r^{(I)} I^{ijR} \partial_R R_{0i0j} + \frac{c_r^{(J)}}{2}J^{b|iRa} \partial_R R_{0iab} \right) \right]\,,\\
&{\rm with}&\qquad c_r^{(I)} = \frac{1}{(r+2)!}\,, \qquad c_r^{(J)} = \frac{2(r+2)}{(r+3)!}\,,
\ee
via the identifications
\be
E=M\,,\quad J^i= L^i\equiv\frac 12 \epsilon^{ijk}L^{j|k}\,,\quad I^{ij}= M^{ij}\,.
\ee
Observe however that usually the radiative momenta appearing in (\ref{sourcelag}) are meant to be traceless, while no such condition applies to $M_{ij}$ in the present appoach.

The energy-momentum moments appearing in (\ref{eq:moments_action}) can be expressed in terms of the binary orbital variables via a matching procedure, which is performed along the lines of \cite{Goldberger:2004jt}, that is by evaluating the {three diagrams of figure \ref{fig:hij} and identifying the appropriate terms.
\begin{figure}
	\begin{center}
		\begin{tikzpicture}
      			\draw [black, thick] (0,0) -- (4.,0);
      			\draw [black, thick] (0,2) -- (4.,2);
			\draw[decorate, decoration=snake, line width=1.5pt, green] (2,2) -- (4,3) ;	
		\end{tikzpicture}\hspace{1cm}
		\begin{tikzpicture}
      			\draw [black, thick] (0,0) -- (4.,0);
      			\draw [black, thick] (0,2) -- (4.,2);
			\draw[decorate, decoration=snake, line width=1.5pt, green] (2,0) -- (4,1) ;	
		\end{tikzpicture}\hspace{1cm}
		\begin{tikzpicture}
      			\draw [black, thick] (0,0) -- (4.,0);
      			\draw [black, thick] (0,2) -- (4.,2);
      			\draw [blue, dashed, line width=1.5] (2,0) -- (2,2);
			\draw[decorate, decoration=snake, line width=1.5pt, green] (2,1) -- (4,1) ;	
		\end{tikzpicture}\hspace{1cm}
	\end{center}
	\caption{Near-zone diagrams relevant for the matching of the  coupling to $h_{ij}$.}
	\label{fig:hij}
\end{figure}
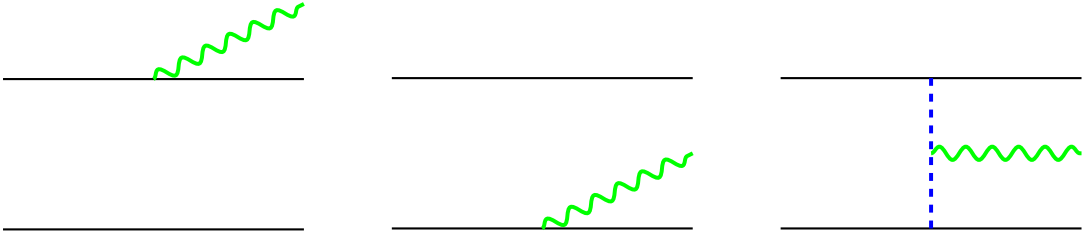

Such procedure results, in the center-of-mass frame and at leading order, in
\be
\label{eq:matching}
\begin{array}{cl}
& M=m\,,\quad M^{ij}=\mu r^i r^j\,,\\
& P^{i,j}=\mu v^i x^j\,,\quad L^{i|j}=\mu\pa{r^i v^j -r^j v^i}=\epsilon_{ijk}L^k\,,\\
& S^{ij}=\mu\pa{v^i v^j -\frac{Gm}{r^3} r^i r^j}\,,
\end{array}
\ee
which clearly comply with the Ward conditions because of the Newtonian dynamics.

More in detail, one could write the identity
\be
S^{ij}=\frac12 \ddot{M}_{ij}-\mu r^{(i}\pa{a^{j)}-a_N^{j)}}\,,\quad a_N^j\equiv -\frac{Gm}{r^3}r^j
\ee
and this is essentially what is done in \cite{Goldberger:2004jt} to set $S^{ij}=\frac12\ddot{M}^{ij}$ at leading order in the action (\ref{eq:moments_action}) and turn it into the more familiar form \ref{sourcelag}.
We will not follow this path here, for two reasons. First: we are interested in the dynamics at NLO in the radiation reaction, so we cannot simply set ${\bf a}={\bf a}_N$. Second: using the eom's in the action is equivalent
to implement a variable shift, which is ok if one ultimately wants to compute gauge invariant quantities, but not in our case, where we also aim at a comparison with the coordinate-dependent MPM results for the radiation reaction and for the memory waveform.

On the other hand, the relation $P^{(i,j)}=\frac12\dot{M}^{ij}$ is an identity which does not rely on the equations of motion, and it can be safely used.
So in the following we will keep working  with (\ref{eq:moments_action}), with $P^{(i,j)}=\frac12\dot{M}^{ij}$, and we will first use it to derive the radiation reaction eom's.
Also, we will neglect $L^{i|j}$ which, at this perturbative order, enters in the so-called {\em failed tails}, which have been already treated elsewere \cite{Almeida:2023yia,Porto:2024cwd}.

\subsection{Radiation reaction}
The leading radiation reaction terms in the equations of motion can be derived from effective action obtained by evaluating the diagrams displayed in Figure \ref{Fig:bubble}.
\begin{figure}
      \begin{tikzpicture}
      \draw [black, thick] (0,0) -- (4.5,0);
      \draw [black, thick] (0,-0.1) -- (4.5,-0.1);
      \draw[decorate, decoration=snake, line width=1.5pt, green] (0.5,0)  arc (180:0:1.75);
      \filldraw[black] (0.5,-0.1) circle (3pt) node[anchor=north] {$S^{ij}_+\,,M^{ij}_+$};
      \filldraw[black] (4.,-0.1) circle (3pt) node[anchor=north] {$S^{ij}_-\,,M^{ij}_-$};
     \draw[-stealth,thick] (1.75,1.25)  arc (150:30:.5);
    \end{tikzpicture}
     \begin{tikzpicture}
      \draw [black, thick] (0,0) -- (4.5,0);
      \draw [black, thick] (0,-0.1) -- (4.5,-0.1);
      \draw[decorate, decoration=snake, line width=1.5pt, green] (0.5,0)  arc (180:0:1.75);
      \filldraw[black] (0.5,-0.1) circle (3pt) node[anchor=north] {$M_+$};
      \filldraw[black] (4.,-0.1) circle (3pt) node[anchor=north] {$S_-\,,I_-$};
     \draw[-stealth,thick] (1.75,1.25)  arc (150:30:.5);
    \end{tikzpicture}
     \begin{tikzpicture}
      \draw [black, thick] (0,0) -- (4.5,0);
      \draw [black, thick] (0,-0.1) -- (4.5,-0.1);
      \draw[decorate, decoration=snake, line width=1.5pt, green] (0.5,0)  arc (180:0:1.75);
      \filldraw[black] (0.5,-0.1) circle (3pt) node[anchor=north] {$S_+\,,I_+$};
      \filldraw[black] (4.,-0.1) circle (3pt) node[anchor=north] {$M_-$};
     \draw[-stealth,thick] (1.75,1.25)  arc (150:30:.5);
    \end{tikzpicture}\hspace{2cm}
     \caption{Bubble diagrams giving the leading order radiation reaction contributions (with the trace of $M_{ij}$ denoted by $I$, the inertia moment)}    
     \label{Fig:bubble}
\end{figure}
Since we are looking for a dissipative effect, we make use of the Keldish formalism and the result is
\be\label{eq:Sbubble}
{\cal S}^{RR}&=&G\int_t\left[\frac3{10}M^{(5)}_{ij+}M^{ij}_- -\frac1{60}I^{(5)}_+ I_- -\frac16\pa{\dddot{S}_+ I_-+\dddot{I}_+ S_-} -2\dot{S}_{ij+}S^{ij}_- + \dot{S}_+ S_-\right.\nn\\
&&\qquad\qquad-\left.\frac16\pa{\dddot{M}_+ I_- + \dddot{I}_+ M_-} -\dot{M}_+ S_- -\dot{S}_+ M_-\right]\,.
\ee
By using the leading order relation $S_{ij}=\frac12 \ddot{M}_{ij}$, the first line would reduce to the form $-\frac G5 Q^{(5)}_{ij+}Q^{ij}_-$  ($Q_{ij}$ being the traceless part of $M_{ij}$), often used in EFT, and whose variation produces the Burke-Thorne acceleration.
We choose however, for the reasons discussed above, not to do this replacement, and we instead compute the eom's by performing directly the variation of (\ref{eq:Sbubble}), obtaining
\be\label{eq:aRR}
a^i_{RR}&\simeq&\ds G\left[\frac35 r^kQ_{ik}^{(5)}+ 2 v^k Q_{ik}^{(4)}-3\frac{Gm}{r^3}r^i\dddot{Q}^{jk}n^j n^k+\frac23 I^{(4)}v^i -\frac43 \dddot{I}\frac{Gm}{r^3}r^i\right]\nn\\
&\simeq&\ds G\left[-\frac25 r^kQ_{ik}^{(5)}+\frac{{\rm d}^2}{{\rm d}t^2}\pa{x^k \dddot{Q}^{ik}+\frac23 \ddot{I} v^i}-\pa{x^k \dddot{Q}^{jk}+\frac23 v^j\ddot{I}}\partial_j a^i_N\right]\,,
\ee
where the $\simeq$ reminds that the result holds up to terms of higher order in the radiation reaction, obtained for instance by replacing an acceleration by ${\bf a}_N$ {\em in the equation of motion}  (not in the action, as we have already stressed).
The first line is the Damour-Deruelle form of the radiation reaction in standard harmonic gauge (SH), while the second line emphasizes the fact that it is related to the Burke-Thorne (BT) form by a world-line shift
\be\label{eq:deltar}
{\bf r}_{BT}={\bf r}_{SH}+\delta {\bf r}\,,\quad \delta r^i= -G \pa{\dddot{Q}^{ik} r^k + \frac 23 v^i \ddot{I}}\,.
\ee
This is the effect,  on the relative coordinate ${\bf r}$, of the well known near zone coordinate transformation imposed on the standard harmonic gauge in order to obtain the radiation reaction in Burke-Thorne form, see page 1001 of \cite{Misner:1973prb}, as well as the generalized treatment done in \cite{Blanchet:2024loi}. It is also worth noticing that also the second line of (\ref{eq:Sbubble}) contributes to the Damour-Deruelle radiation reaction at this order, more precisely the terms proportional to $M_-$, while the other terms containing $M_+$ affect the dynamics only at 5PN.

\section{Quadrupole-quadrupole amplitudes}
\label{sec:WF}
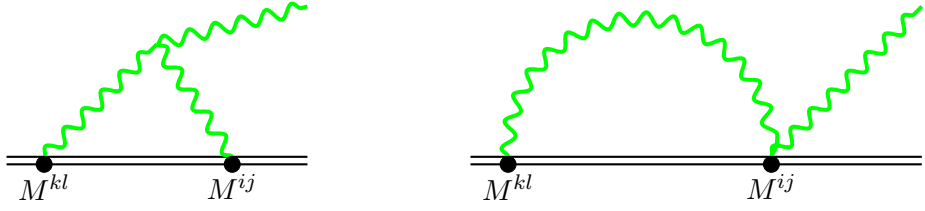
\begin{figure}
  \begin{center}
    \begin{tikzpicture}
      			\draw [black, thick] (0,0) -- (4.,0);
      			\draw [black, thick] (0,-0.1) -- (4.,-0.1);
      			\draw [decorate, decoration=snake, line width=1.5pt, green] (3,0) -- (2,1.5);
			\draw[decorate, decoration=snake, line width=1.5pt, green] (.5,0) -- (2,1.5);
			\draw[decorate, decoration=snake, line width=1.5pt, green] (2,1.5) -- (4,2) ;
      			\filldraw[black] (3.,-0.1) circle (3pt) node[anchor=north] {$M^{ij}$};
      			\filldraw[black] (.5,-0.1) circle (3pt) node[anchor=north] {$M^{kl}$};
		\end{tikzpicture}\hspace{2cm}
    	\begin{tikzpicture}
      		\draw [black, thick] (0,0) -- (6,0);
      		\draw [black, thick] (0,-0.1) -- (6,-0.1);
      		\draw[decorate, decoration=snake, line width=1.5pt, green] (0.5,0)  arc (180:0:1.75);
      		\draw[decorate, decoration=snake, line width=1.5pt, green] (4,0) -- (6,2);
      		\filldraw[black] (0.5,-0.1) circle (3pt) node[anchor=north] {$M^{kl}$};
      		\filldraw[black] (4.,-0.1) circle (3pt) node[anchor=north] {$M^{ij}$};
    	\end{tikzpicture}
    \caption{The two NLO diagrams giving quadrupole-quadrupole terms in the emission amplitude. On the left, the ``pure memory" diagram; on the right, the contact diagram.}    
    \label{fig:Ade}
  \end{center}
\end{figure}
The NLO contributions to the waveform come, as well as from the NLO evaluation of the simple emission amplitudes, from the more complex diagrams displayed in Figure \ref{fig:Ade}.
It is worth stressing that, at this perturbative order, the standard EFT replacement $S_{ij}=\frac12 \ddot{M}_{ij}$ can be safely made already at the diagramatic level because any difference would show further up in the perturbative scheme. This means that, to all purposes, one can practically compute the diagrams using the linear coupling action in the form (\ref{sourcelag}); this will be particularly useful later on in this section.
We stress again that at leading order we still have to rely on the amplitude in the form
${\cal A}^{\rm LO}_{ij}\simeq\frac12S_{ij}$, which contains an ${\cal O}\pa{G}$ term when expressed in terms of the mass quadrupole.

\subsection{``Pure memory" diagram}
\label{sec:mem_traces}
We start from the left diagram in figure \ref{fig:Ade} (and we use the matching condition $I_{ij}=M_{ij}$).
The calculation is long but straightforward and we report here the onshell ($\K^2=\omega^2$) result for all the amplitude components (with $\omega_1\equiv\omega-\omega'$):
\be
\ds i \mathcal{A}^{\rm mem}_{ij} (\omega,\omega{\bf n})=-\frac{G}4\int_{\omega'}
M_{kl}(\omega_1)M_{mn}(\omega')
\left\{\delta_{k(i}\delta_{j)m}\delta_{ln}\omega'^3\pa{-\frac{13{\omega'}^3}{45\omega}+\frac 13\omega_1\omega'  +2\omega_1^2}\right.\nonumber\\
\qquad
+n^kn^m\delta^{l(i}\delta^{j)n}\omega'^3\pa{\frac 1{45}\frac{{\omega'}^3}\omega+  \frac 4{15}{\omega'}^2  +\omega_1\omega'  +\frac23 \omega_1^2}\nonumber\\
\qquad +\frac 1{90}n^kn^l\delta^{mi}\delta^{jn}\pa{\frac{{\omega'}^6}{\omega}  +4{\omega'}^5 -7{\omega'}^4\omega_1+  \frac{89}2{\omega'}^3\omega_1^2+\frac{91}{2}{\omega'}^2\omega_1^3  +\frac {29}{2}{\omega'}\omega_1^4+\frac12\omega_1^5}\nonumber\\
\qquad +\frac1{90}\delta_{mn}\delta_{k(i}\delta_{j)l}\pa{17\frac{{\omega'}^6}{\omega}  -\frac{17}2{\omega'}^5 -\frac{13}2{\omega'}^4\omega_1- \frac{287}2{\omega'}^3\omega_1^2+\frac{107}{2}{\omega'}^2\omega_1^3 +44{\omega'}\omega_1^4+31\omega_1^5}\nonumber\\
\qquad +\delta_{ij}\left[n_k n_l n_m n_n\frac{\omega'^3}6\pa{\frac 1{10}\frac{{\omega'}^3}\omega-\frac{11}{20}{\omega'}^2-\frac54\omega_1\omega'-\omega_1^2}\right.\nonumber\\
\qquad\quad
+\frac1{45}n_n n_m \delta_{kl}\pa{4\frac{{\omega'}^6}{\omega}  -\frac{61}8{\omega'}^5+\frac18{\omega'}^4\omega_1-2{\omega'}^3\omega_1^2 +2{\omega'}^2\omega_1^3 +\frac{59}8{\omega'}\omega_1^4+\frac{25}8\omega_1^5}\nn\\
\qquad\quad +n_m n_k \delta_{ln}\omega'^3\pa{-\frac 7{45}\frac{{\omega'}^3}\omega+  \frac7{15}{\omega'}^2+\frac23\omega_1\omega'+\frac23\omega_1^2}\nn\\
\qquad\quad +\delta_{km}\delta_{ln}\omega'^3\pa{\frac{31}{90}\frac{{\omega'}^3}\omega- \frac15{\omega'}^2 +\frac56\omega_1\omega'}\nn\\
\qquad\quad\left.+\frac1{160}\delta_{kl}\delta_{mn}\omega'^3\pa{-\frac{232}{9}\frac{{\omega'}^3}\omega+\frac{276}9{\omega'}^2-20\omega_1\omega'
  +\frac{80}3\omega_1^2}\right]\nn\\
\qquad +n_i n_j\left[n_k n_l n_m n_n\omega'^3\pa{-\frac{49}{40}\frac{{\omega'}^3}\omega+ \frac{73}{60}{\omega'}^2-\frac{31}{24}\omega_1\omega'+\frac{11}{12}\omega_1^2}\right.\nonumber\\
\qquad\quad+n_n n_m \delta_{kl}\pa{-\frac{41}{20}\frac{{\omega'}^6}{\omega}  +\frac{49}{24}{\omega'}^5 -\frac{31}{15}{\omega'}^4\omega_1+\frac{233}{120}{\omega'}^3\omega_1^2-\frac{73}{120}{\omega'}^2\omega_1^3 -\frac4{15}{\omega'}\omega_1^4-\frac7{120}\omega_1^5}\nn\\
\qquad\quad+n_m n_k \delta_{ln}\omega'^3\pa{\frac{127}{30}\frac{{\omega'}^3}\omega-  \frac{21}5{\omega'}^2+\frac92\omega_1\omega'-3\omega_1^2}\nn\\
\qquad\quad +\delta_{km}\delta_{ln}\omega'^3\pa{-\frac{299}{180}\frac{{\omega'}^3}\omega+\frac{33}{20}{\omega'}^2-\frac{11}6\omega_1\omega'+\omega_1^2}\nn\\
\qquad\quad\left.+\frac1{12}\delta_{kl}\delta_{mn}\omega'^3\pa{\frac{301}{30}\frac{{\omega'}^3}\omega-\frac{53}{5}{\omega'}^2+\frac{19}2\omega_1\omega'-7\omega_1^2}\right]\nn\\
\qquad+\frac1{15}n_k n_{(i}\left[\delta_{j)l}n_n n_m \pa{2\frac{{\omega'}^6}{\omega}  -2{\omega'}^5 -\frac14{\omega'}^4\omega_1-11{\omega'}^3\omega_1^2+\frac{27}{2}{\omega'}^2\omega_1^3 +9{\omega'}\omega_1^4+\frac94\omega_1^5}\right.\nn\\
\qquad\quad+\frac13\delta_{j)m}\delta_{ln}\pa{-28\frac{{\omega'}^6}{\omega}  -13{\omega'}^5 -\frac{283}2{\omega'}^4\omega_1-76{\omega'}^3\omega_1^2+31{\omega'}^2\omega_1^3 -{\omega'}\omega_1^4-\frac12\omega_1^5}\nn\\
\qquad\quad\left.\left.+\frac13\delta_{j)l}\delta_{mn}\pa{16\frac{{\omega'}^6}{\omega}  -\frac{13}2{\omega'}^5 +\frac{167}4{\omega'}^4\omega_1+7{\omega'}^3\omega_1^2+\frac12{\omega'}^2\omega_1^3+\frac{29}2{\omega'}\omega_1^4+\frac{29}4\omega_1^5}\right]\right\}\,,\nn\\
\ee

\be
\ds i \mathcal{A}^{\rm mem}_{0i}(\omega,\omega{\bf n})=\frac G8\int_{\omega'}
M_{kl}(\omega_1)M_{mn}(\omega')
\left\{n_{i}\left[\delta_{mk}\delta_{ln}\omega'^3\pa{-\frac{79}{30}\frac{{\omega'}^3}{\omega}+\frac{29}{10}\omega'^2-2\omega_1\omega'  +2\omega_1^2}\right.\right.\nonumber\\
\qquad\quad \delta_{mn}\delta_{kl}\omega'^3\pa{\frac{27}{20}\frac{{\omega'}^3}{\omega}-\frac{83}{60}\omega'^2+\frac43\omega_1\omega'  -\frac56\omega_1^2}\nonumber\\
\qquad\quad
+n_l n_n\delta_{mk}\omega'^3\pa{\frac{113}{15}\frac{{\omega'}^3}{\omega}-\frac{233}{30}\omega'^2+\frac{43}6\omega_1\omega'  -\frac{17}3\omega_1^2}\nonumber\\
\qquad \quad+n_l n_k n_m n_n\omega'^3\pa{-\frac{137}{60}\frac{{\omega'}^3}{\omega}+\frac{34}{15}\omega'^2-\frac{29}{12}\omega_1\omega'  +\frac{5}3\omega_1^2}
\nonumber\\
\qquad \quad\left.+\frac1{10}n_m n_n\delta_{kl}\pa{-\frac{107}{3}\frac{{\omega'}^6}{\omega}  +\frac{71}2{\omega'}^5 -\frac{69}2{\omega'}^4\omega_1+  \frac{69}2{\omega'}^3\omega_1^2-\frac{37}{6}{\omega'}^2\omega_1^3  +\frac{11}3{\omega'}\omega_1^4+\frac73\omega_1^5}\right]\nonumber\\
\qquad+\frac15\delta_{mi}\left[n_k \delta_{nl}\pa{-6\frac{{\omega'}^6}{\omega}  +12{\omega'}^5 +\frac{9}2{\omega'}^4\omega_1+  3{\omega'}^3\omega_1^2+12{\omega'}^2\omega_1^3  +3{\omega'}\omega_1^4-\frac32\omega_1^5}\right.\nn\\
\qquad\quad n_n \delta_{kl}\pa{\frac{11}3\frac{{\omega'}^6}{\omega}  +\frac7{12}{\omega'}^5 +\frac{61}6{\omega'}^4\omega_1+  \frac73{\omega'}^3\omega_1^2+\frac{91}6{\omega'}^2\omega_1^3  +\frac{83}{12}{\omega'}\omega_1^4-2\omega_1^5}\nn\\
\qquad\quad \left.\left.n_n n_k n_l\pa{\frac{{\omega'}^6}{\omega}  +\frac7{4}{\omega'}^5 +8{\omega'}^4\omega_1+  12{\omega'}^3\omega_1^2+\frac{11}2{\omega'}^2\omega_1^3  +\frac{23}{4}{\omega'}\omega_1^4+\frac32\omega_1^5}
\right]\right\}\,,\nn\\
\ee
\be
\ds i \mathcal{A}^{\rm mem}_{00}(\omega,\omega{\bf n})=G\int_{\omega'}
M_{kl}(\omega_1)M_{mn}(\omega')
\left\{\delta_{mk}\delta_{ln}\omega'^3\pa{\frac{11}{48}\frac{{\omega'}^3}{\omega}-\frac{67}{240}\omega'^2-\frac16\omega_1\omega'  +\frac14\omega_1^2}\right.\nonumber\\
\qquad\quad \delta_{mn}\delta_{kl}\omega'^3\pa{-\frac{13}{96}\frac{{\omega'}^3}{\omega}+\frac{43}{480}\omega'^2-\frac1{12}\omega_1\omega'  -\frac{19}{48}\omega_1^2}\nonumber\\
\qquad\quad
+n_l n_n\delta_{mk}\omega'^3\pa{-\frac{19}{24}\frac{{\omega'}^3}{\omega}+\frac{17}{24}\omega'^2-\frac{13}{12}\omega_1\omega'  +\frac13\omega_1^2}\nonumber\\
\qquad \quad+n_l n_k n_m n_n\omega'^3\pa{\frac{25}{96}\frac{{\omega'}^3}{\omega}-\frac{35}{96}\omega'^2-\frac{1}{24}\omega_1\omega'  -\frac{31}{48}\omega_1^2}
\nonumber\\
\qquad \quad\left.+n_m n_n\delta_{kl}\pa{\frac{17}{48}\frac{{\omega'}^6}{\omega}  -\frac{11}{24}{\omega'}^5 +\frac{17}{96}{\omega'}^4\omega_1- \frac{47}{96}{\omega'}^3\omega_1^2-\frac{29}{96}{\omega'}^2\omega_1^3  -\frac{37}{96}{\omega'}\omega_1^4-\frac7{48}\omega_1^5}
\right\}\,.\nonumber\\
\ee

There is an additional, purely longitudinal term proportional to the integral
$I_m(\omega_1,\omega')\equiv\int_{\Q}\paq{(\K-\Q)^2-\omega_1^2}^{-1}\paq{\Q^2-{\omega'}^2}^{-1}$:
\be
\label{eq:Im}
\mathcal{A}^{\rm mem}_{L\mu\nu}(\omega,\K)\equiv -i \pi G\int_{\omega'}I_m
M_{kl}(\omega_1)M_{mn}(\omega')
\omega'^3\omega_1^3\Lambda^{TT}_{kl,mn} \frac{k^{\mu}k^{\nu}}{\omega^2}\,,
\ee
where $\Lambda^{TT}$ is the standard tranverse traceless (TT) projector
depending on ${\bf n}\equiv{\bf \hat k}$.
This term trivially satisfies the Ward identities
and it does not contribute to the TT part of the amplitude.

\subsection{Contact diagram}
For computing the diagram on the right in figure \ref{fig:Ade},
one needs to extract the source
quadratic coupling to the gravitational field.
Moreover, and related to this, equation (\ref{sourcelag}) needs to be made covariant.
When evaluated in the rest frame of an observer at infinity (such that $u^\mu\equiv\pa{u^0,0,0,0}$, with $u^2=-1$), this brings to the following coupling terms\footnote{This results in a different Lagrangian than the one used in \cite{Porto:2024cwd}, where the multipoles are defined in a locally free-falling frame, which is connected to the covariant metric $g_{\mu\nu}$ by means of a tetrad field. Here we follow instead the MPM conventions for the frame choice, which do not require the introduction of tetrad fields. The other obvious difference, as already pointed out, is that our multipoles are not traceless.}:
\be\label{eq:quadratic}
    d\tau I^{ij} \pa{R_{0 i 0 j}u^0 u^0}= dt\pa{1-\frac{h_{00}}2} I^{ij}R_{0i0j}\pa{1+h_{00}}+{\cal O}\pa{h^3} \,.
\ee
Having so established the Feynman's rule for the quadratic matter interaction vertex, the contact diagram can be evaluated to
\be\label{eq:contactwf}
&&\left.\begin{array}{c}
i \mathcal{A}^{\rm c}_{00}(\omega,\K)\\
i \mathcal{A}^{\rm c}_{oa}(\omega,\K)\\
i \mathcal{A}^{\rm c}_{ab}(\omega,\K)
\end{array}\right\}
=-\frac{G}2 \int \frac{{\rm d}\omega'}{2\pi} M^{ij}(\omega_1) M^{kl}(\omega') \omega'^3\times\\
&&\qquad\times\left\{\begin{array}{l}
\ds \frac{1}{30} \Big[\delta_{jl} \delta_{ik}(30 \omega_1^2 -25 \omega_1 \omega' - \omega'^2)-\delta_{kl}\delta_{ij}\pa{30 \omega_1^2 -5 \omega_1\omega'+3 {\omega'}^2} -10 \delta_{kl}k_i k_j\Big]\\
\ds-\frac12 \Big[\delta_{ik}\omega'  (k_l \delta_{ja} - k_j \delta_{la})-\frac23 \delta_{kl}\delta_{ai}k_j \omega_1\Big]\\
\ds \Big[\omega \omega' \delta_{ik}\delta_{j(a} \delta_{b)l}-\frac13\delta_{kl}\delta_{ia} \delta_{jb}\pa{-3\omega^2_1+\omega'^2}\Big]\,.
\end{array}\right.\nn
\ee
This contribution has no counterpart in the MPM derivation, where just the analog of the ``pure memory" diagram is evaluated at first, as a particular solution of the linearized Einstein's equations
\be
\Box\bar{h}_{\mu\nu}=8 \pi G \pa{T_{\mu\nu}+\tau_{\mu\nu}}\,.
\ee
In the MPM formalism, then a solution of the homogeneous equation $\Box\bar{h}_{\mu\nu}=0$ is added in such a way to obtain a waveform with the correct boundary conditions.
Such homogeneous solution corresponds precisely to the term reported here in (\ref{eq:contactwf}), and it is quite satisfying that it can be derived from first principles in the EFT formalism.

\subsection{Ward identities}
The timelike part gives
\be
  \label{eq:wmem0_rik}
  \ds ik^\mu \pa{\mathcal{A}^{\rm LO}_{\mu 0}+\mathcal{A}^{\rm mem}_{\mu 0}+\mathcal{A}^{\rm c}_{\mu 0}}
&=&\ds\paq{i\frac{\omega}2 M(\omega)+\frac G{10}\int_{\omega'} \omega'^5 \omega_1 Q^{kl}(\omega_1)Q^{kl}(\omega')}\,,
\ee
while the spacelike part reads
\be
\label{eq:wmemi_rik}
&&\ds ik^\mu \pa{\mathcal{A}^{\rm LO}_{\mu i}+\mathcal{A}^{\rm mem}_{\mu i}+\mathcal{A}^{\rm c}_{\mu i}}
=\ds k_l\Big\{\frac 14\omega L_{i|l}(\omega)+
      \frac {G}{10}\int_{\omega'}\pa{\omega_1^5-{\omega'}^5}Q^{lk}(\omega_1)Q^{ki}(\omega')\\
&&\ds\qquad\qquad\qquad +\frac i2\bigg[S_{il}-\frac 12\ddot M_{il} +i\frac G5\int_{\omega'} \bigg(\pa{{\omega'}^5+\omega_1^5}Q^{kl}(\omega_1)Q^{ki}(\omega')+\frac 23I(\omega_1)Q^{il}(\omega'){\omega'}^5\bigg) \bigg]\Big\} \nn\,.
\ee
In time domain this implies
\be
\label{eq:Wards}
&&\dot{M}=-\frac G5\dot{Q}_{jk}Q^{(5)}_{jk}\,,\quad \dot{L}_{i|j}=\frac45GQ_{k[i}^{(5)}Q_{j]k}\\
&&S_{ij}=\frac12\ddot{M}_{ij}+\frac25GQ_{k(i}^{(5)}M_{j)k}\,.
\ee
Notice that, if we take the expressions for $M$, $L^{i|j}$ and $S_{ij}$ (adding to $M$ the newtonian part of the energy) given by the {\em leading order} matching procedure, eqns.(\ref{eq:matching}), such conditions are {\em not} compatible with the equations of motion (\ref{eq:aRR}). This indicates that the matching (\ref{eq:matching}) is not the end of the story, and must be complemented with a 2.5PN relative part, say $M_{RR}$, $L^{i|j}_{RR}$, $S^{ij}_{RR}$ in order to reconcile the eom's with the Ward identities.
Also, one remarks that the latter would be identically satisfied by (\ref{eq:matching}) if the equations of motion were in the Burke-Thorne form. This is not surprising, as the Burke-Thorne frame is known to be the only one where the RR equations of motion are not "contaminated" by interference with Newtonian contributions, like the third term in (\ref{eq:aRR}), see \cite{Blanchet:2024loi} and references therein for more details.
So, in some sense, it is a privileged frame which is indeed universally adopted in EFT treatment (and also very often employed in the other approaches); we did not follow the same route here because we aim to investigate the relationship between the two frames from an EFT point of view. The radiation reaction force in the form (\ref{eq:aRR}) induces Schott terms in
the energy and angularm momentum derivatives, as reported in App.~\ref{app:Schott}.

\subsection{TT waveform in direct space}
To find the transverse-traceless part of the gravitational perturbation in
direct space, we apply (\ref{eq:hexpect}) to $\pa{\mathcal{A}^{\rm LO}+\mathcal{A}^{\rm mem}+\mathcal{A}^{\rm c}}_{ij}$ and we integrate using the following formula valid in $d=3$:
\begin{align}
\int_\K \frac{e^{i\K\cdot\X}}{(\K^2-\omega^2)} k^{i_1} \dots k^{i_m} &= \frac{e^{i\omega R}}{4\pi R}\omega^m n^{i_1}\dots n^{i_m} + {\cal O}\pa{\frac1{r^2}}\,.
\end{align}
The result is
\be
h_{ij}^{TT}&=&\frac{4 G}R\Lambda_{ij,kl}\left\{S^{kl}-\frac G{90}\left[\pa{77 M_{b(k}^{(5)}M_{l)b}+118 M_{b(k}^{(4)}\dot{M}_{l)b}+77 \dddot{M}_{b(k}\ddot{M}_{l)b}}
\right.\right.\\
&&\qquad\qquad+n^a n^b \pa{13M_{a(k}^{(5)}M_{l)b}+44M_{a(k}^{(4)}\dot{M}_{l)b}+31\dddot{M}_{a(k}\ddot{M}_{l)b}}\nn\\
&&+\frac{n^a n^b}4\pa{10M_{kl}^{(5)}M_{ab}-16M_{kl}^{(4)}\dot{M}_{ab}+91\dddot{M}_{kl}\ddot{M}_{ab}+91\ddot{M}_{kl}\dddot{M}_{ab}+29\dot{M}_{kl}M^{(4)}_{ab}+M_{kl}M^{(5)}_{ab}}\nn\\
&&+\left.\frac14\pa{62M_{kl}^{(5)}I+88M_{kl}^{(4)}\dot{I}+107\dddot{M}_{kl}\ddot{I}+107\ddot{M}_{kl}\dddot{I}-47\dot{M}_{kl}I^{(4)}-103M_{kl}I^{(5)}}\right]\nn\\
&&\qquad\qquad\left.-\frac 12\int_t \paq{\frac{13}{45}\dddot{M}_{ak}\dddot{M}_{la}-\frac1{45}n^a n^b\dddot{M}_{ak}\dddot{M}_{lb}-\frac1{90}n^a n^b\dddot{M}_{kl}\dddot{M}_{ab}-\frac{17}{90}\dddot{M}_{kl}\dddot{M}_{bb}}\right\}\nn\\
\label{eq:hTT}
&=&\frac{4 G}R\Lambda_{ij,kl}\left\{\frac12\ddot{Q}^{kl}-\frac G{90}\left[\pa{41 Q_{b(k}^{(5)}Q_{l)b}+118 Q_{b(k}^{(4)}\dot{Q}_{l)b}+77 \dddot{Q}_{b(k}\ddot{Q}_{l)b}}
\right.\right.\\
&&\qquad\qquad+n^a n^b \pa{13Q_{a(k}^{(5)}Q_{l)b}+44Q_{a(k}^{(4)}\dot{Q}_{l)b}+31\dddot{Q}_{a(k}\ddot{Q}_{l)b}}\nn\\
&&+\frac{n^a n^b}4\pa{10Q_{kl}^{(5)}Q_{ab}-16Q_{kl}^{(4)}\dot{Q}_{ab}+91\dddot{Q}_{kl}\ddot{Q}_{ab}+91\ddot{Q}_{kl}\dddot{Q}_{ab}+29\dot{Q}_{kl}Q^{(4)}_{ab}+Q_{kl}Q^{(5)}_{ab}}\nn\\
&&+\left.30\pa{Q_{kl}^{(5)}I+2Q_{kl}^{(4)}\dot{I}+2\dddot{Q}_{kl}\ddot{I}+2\ddot{Q}_{kl}\dddot{I}+\dot{Q}_{kl}I^{(4)}}\right]\nn\\
&&\qquad\qquad\left.-\frac 12\int_t \paq{\frac{13}{45}\dddot{Q}_{ak}\dddot{Q}_{la}-\frac1{45}n^a n^b\dddot{Q}_{ak}\dddot{Q}_{lb}-\frac1{90}n^a n^b\dddot{Q}_{kl}\dddot{Q}_{ab}}\right\}\,.\nn
\ee
Notice the crucial replacement of $S_{kl}$ with $\frac12\ddot{Q}_{kl}+\frac25 G M_{j(k}Q^{(5)}_{l)j}$ after the second equality as dictated by the Ward condition.

\subsection{Comparison with the literature}
Comparison with \cite{Blanchet:1997ji} shows perfect agreement, modulo what might look as a redefinition of the quadrupole moment:
\begin{equation}\label{eq:Qshift}
Q_{ij} \rightarrow Q_{ij} + \frac{8G}{7} \dddot{Q}_{k<i} Q_{j>k} + \frac{2G}{3} (\dddot{I} Q_{ij} +  \dddot{Q}_{ij} I)\,,
\end{equation}
where $<ij>$ denotes symmetrization and trace removal.
However the need of this redefinition is just apparent, because the quadrupole in the MPM formalism contains a 2.5PN contribution, see for instance
Eq. (3.1) of \cite{Arun:2007sg} when written in terms of orbital variables. A more general expression for the same term can be derived from eqs.(4.12) and (4.24) of \cite{Blanchet:1996wx}, which are however incomplete because of a ``hidden'' contribution coming from the definition of $\sigma$, contained in (4.3) of the same paper\footnote{We thank Luc Blanchet for clarifying this important point to us.}.
Once the latter ``hidden'' contribution is taken into account, the sign of the first term of (4.12) of \cite{Blanchet:1996wx} is swapped, and the resulting total 2.5PN contribution to the MPM quadrupole can be shown to match exactly the expression reported above in (\ref{eq:Qshift}), as well as eq. (3.1) of \cite{Arun:2007sg}.
We thus conclude that we have reproduced the
quadrupole-quadrupole waveform reported in \cite{Blanchet:1997ji}.

Coming to \cite{Porto:2024cwd}, the waveform is not explicitly reported there, but is it possible to reconstruct it and to track down the differences with our result, which stem from some different choices made along the derivation. First, the usual EFT action (\ref{sourcelag}), with traceless quadrupoles, is used there; this implies, among other things, the use of the Burke-Thorne form for the RR equations of motion. Second, a vierbein is introduced in the covariantization procedure, resulting in a different  quadratic coupling to the multipole source. As a result, the waveform derived in \cite{Porto:2024cwd}, which we denote by $h^{TT}_{BT}$, is different from (\ref{eq:hTT}):

\be\label{eq:comphTT}
h^{TT}_{BT}=h^{TT}_{(\ref{eq:hTT})}+\frac{4 G^2}R\Lambda_{ij,kl}\frac12\ddot{\delta Q}_{kl}\,,\quad \delta Q_{kl}\equiv \dddot{Q}_{b(k}M_{l)b}+\frac13 \dot{Q}_{kl}\ddot{I}\,.
\ee
The term $\delta Q_{kl}$ can be recognized as the variation induced in the second mass moment $M_{kl}=\mu r^k r^l$ due to the shift $\delta {\bf r}$ written in (\ref{eq:deltar}).
So, if one takes into account that the two derivations are done in different frames or, in other words, are written using different variables to describe the relative coordinate ${\bf r}$, the two waveforms turn out to be equivalent.

\section{Effective action}
\label{sec:Sinin}
In this section, we derive the in-in effective action associated with the quadrupole-quadrupole emission, consistent with the picture presented above, and we show that it provides energy and angular momentum losses compatible with the gravitational-wave fluxes at infinity.
\begin{figure}
  \begin{center}
        \begin{tikzpicture}
      \draw [black, thick] (0,0) -- (4.5,0);
      \draw [black, thick] (0,-0.1) -- (4.5,-0.1);
      \draw[decorate, decoration=snake, line width=1.5pt, green] (0.5,0)  arc (180:0:1.75);
      \filldraw[black] (0.5,-0.1) circle (3pt) node[anchor=north] {$+$};
      \filldraw[black] (4.,-0.1) circle (3pt) node[anchor=north] {$-$};
     \draw[-stealth,thick] (1.75,1.25)  arc (150:30:.5);
    \end{tikzpicture}\hspace{2cm}
    \begin{tikzpicture}
      \draw [black, thick] (0,0) -- (4.5,0);
      \draw [black, thick] (0,-0.1) -- (4.5,-0.1);
      \draw[decorate, decoration=snake, line width=1.5pt, green] (2.25,0) -- (2.25,1.85) ;
      \draw[decorate, decoration=snake, line width=1.5pt, green] (0.5,0)  arc (180:0:1.75);
      \filldraw[black] (0.5,-0.1) circle (3pt) node[anchor=north] {$+$};
      \filldraw[black] (2.25,-0.1) circle (3pt) node[anchor=north] {$+$};
      \filldraw[black] (4.,-0.1) circle (3pt) node[anchor=north] {$-$};
       \draw[-stealth,thick] (2.5,.4)--(2.5,1.1);
      \draw[-stealth,thick] (1.2,.9)  arc (180:90:.5);
      \draw[-stealth,thick] (2.7,1.4)  arc (90:0:.5);
    \end{tikzpicture}\\
      \begin{tikzpicture}
      \draw [black, thick] (0,0) -- (8,0);
      \draw [black, thick] (0,-0.1) -- (8,-0.1);
      \draw[decorate, decoration=snake, line width=1.5pt, green] (0.5,0)  arc (180:0:1.75);
      \draw[decorate, decoration=snake, line width=1.5pt, green] (4.,0)  arc (180:0:1.75);
      \filldraw[black] (0.5,-0.1) circle (3pt) node[anchor=north] {$+$};
      \filldraw[black] (4.,-0.1) circle (3pt) node[anchor=north] {$-$};
      \filldraw[black] (7.5,-0.1) circle (3pt) node[anchor=north] {$+$};
           \draw[-stealth,thick] (1.75,1.25)  arc (150:30:.5);
           \draw[stealth-,thick] (5.25,1.25)  arc (150:30:.5);
    \end{tikzpicture}
    \begin{tikzpicture}
      \draw [black, thick] (0,0) -- (8,0);
      \draw [black, thick] (0,-0.1) -- (8,-0.1);
      \draw[decorate, decoration=snake, line width=1.5pt, green] (0.5,0)  arc (180:0:1.75);
      \draw[decorate, decoration=snake, line width=1.5pt, green] (4.,0)  arc (180:0:1.75);
      \filldraw[black] (0.5,-0.1) circle (3pt) node[anchor=north] {$+$};
      \filldraw[black] (4.,-0.1) circle (3pt) node[anchor=north] {$+$};
      \filldraw[black] (7.5,-0.1) circle (3pt) node[anchor=north] {$-$};
           \draw[-stealth,thick] (1.75,1.25)  arc (150:30:.5);
           \draw[-stealth,thick] (5.25,1.25)  arc (150:30:.5);
    \end{tikzpicture}
    \caption{From top to bottom, left to right: leading order, pure memory, and double contact diagram(s) using Keldish variables and oriented propagators.}
    \label{fig:QQQ}
  \end{center}
\end{figure}
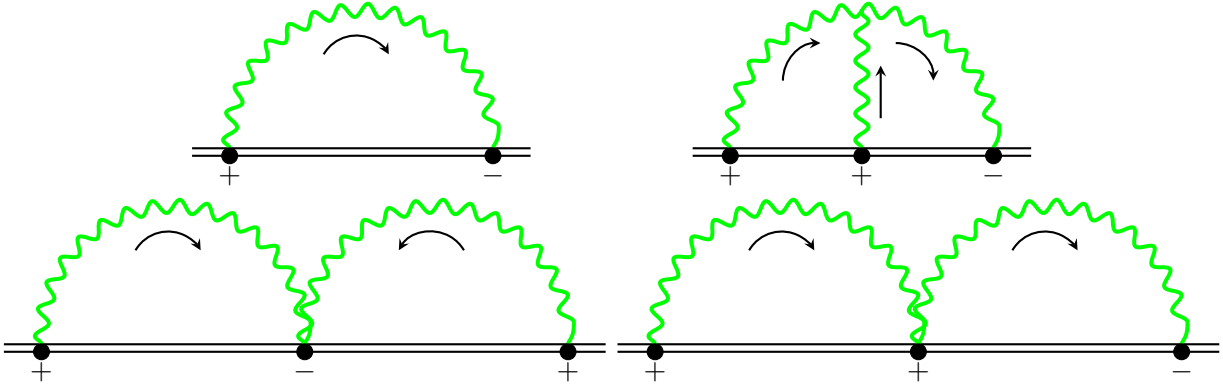

\subsection{Diagrams}
We need to evaluate the diagrams displayed in figure \ref{fig:QQQ}.
The simplest diagram (top left)
have already been computed, see (\ref{eq:Sbubble}), and studied at leading order (2.5PN).
Here we aim to extend the analysis at next to leading order in the radiation reaction (5PN),  which is the same order of the other diagrams appearing in figure \ref{fig:QQQ}..

To do so, we leave the auto-imposed constraint adopted in section \ref{sec:setup}, that is not using the equations of motion in the effective action: now we have understood that it corresponds to a change of coordinate from the Burke-Thorne to the Damour-Deruellle form of the radiation reaction equation, there is no need of unnecessary complication. So we allow ourselves to use the equations of motion in the action, namely in the form dictated by the Ward identities (\ref{eq:Wards}).

 At leading order we get, as preiously mentioned, the usual
quadrupole-quadrupole effective action responsible for the Burke-Thorne
radiation reaction term involving traceless quadrupoles
  \be
  \label{eq:QQ}
{\cal S}_{Q^2}&=&-\frac G5\int_t Q^{ij}_-Q^{(5)}_{ij\,,+}\,.
\ee
At 5PN one has
\be
{\cal S}_{QQ,NLO}&=&-\frac{2G^2}5\int_t\left\{\paq{-\dddot{Q}^{ij}_- M^{ki}_+ Q^{(5)}_{kj,+}+\dddot{Q}^{ij}_+\pa{M^{ki}_-Q^{(5)}_{kj,+}+M^{ki}_+Q^{(5)}_{kj,-}}}\right.\\
&&\qquad\qquad\qquad\qquad\left.+\frac13\paq{\pa{\dot{Q}^{kl}_+Q^{(5)}_{kl,-}+\dot{Q}^{kl}_-Q^{(5)}_{kl,+}}\ddot{I}_+ -\dot{Q}^{kl}_+Q^{(5)}_{kl,+}\ddot{I}_-}\right\}\nn\\
&=&-\frac{2G^2}5\int_t\left\{\paq{2Q_{+ij}^{(4)}\dddot{Q}_{-jk}\dot{M}_{+}^{ik}-Q_{+ij}^{(4)}Q_{+jk}^{(4)}M_{-}^{ik}+\dddot{Q}_{+ij}\dddot{Q}_{-jk}\ddot{M}_{+}^{ik}+\frac12\dddot{Q}_{+ij}\dddot{Q}_{+jk}\ddot{M}_{-}^{ik}}\right.\nn\,.\\
&&\qquad\qquad\qquad-\frac13 \left[\dddot{Q}^{kl}_+\dddot{Q}_{kl,+} \ddot{I}_-+\pa{\dddot{Q}^{kl}\ddot{Q}_{kl}-Q^{(4)}_{kl}\dot{Q}^{kl}}_+\dddot{I}_-\right.\nn\\
&&\qquad\qquad\qquad\quad\left.\left.-2\dddot{Q}^{kl}_+\dddot{Q}_{kl,-} \ddot{I}_+-\pa{\dddot{Q}^{kl}_+\dot{Q}_{kl,-}-2\ddot{Q}^{kl}_-\ddot{Q}_{kl,+}+\dddot{Q}^{kl}_-\dot{Q}_{kl,+}}I^{(4)}_+\right]\right\}\,.
\ee 

The top right diagram in figure \ref{fig:QQQ} has been evaluated in \cite{Blumlein:2021txe} and \cite{Almeida:2022jrv} for the case of traceless quadrupoles. We complete here the result with the inclusion of traces:
\be
\label{eq:actmem}
{\cal S}_{\rm bulk}&=&G^2\int_t\left\{\frac15Q_{+ij}^{(4)}Q_{+jk}^{(4)}Q_{-}^{ik}-\frac25Q_{+ij}^{(4)}Q_{+jk}\pa{Q_{-}^{ik}}^{(4)}+\frac8{35}\dddot{Q}_{+ij}\dddot{Q}_{+jk}\ddot{Q}_{-}^{ik}-\frac{12}{35}\dddot{Q}_{+ij}\ddot{Q}_{+jk}\dddot{Q}_{-}^{ik}\right.\nn\\
&&-\frac1{15}I_+Q_{+ik}^{(4)}\pa{Q_-^{ik}}^{(4)}-\frac2{15}\ddot{I}_+\dddot{Q}_{+ik}\dddot{Q}_-^{ik}+\frac{14}{15}\dddot{I}_+\pa{\ddot{Q}_{+ik}\dddot{Q}_{-}^{ik}-\dddot{Q}_{+ik}\ddot{Q}_-^{ik}}\nn\\
&&+I_+^{(4)}\paq{\frac1{45}\pa{Q_{+ik}\pa{Q_-^{ik}}^{(4)}-Q_{+ik}^{(4)}Q_{-}^{ik}}+\frac25\pa{\dot{Q}_{+ik}\dddot{Q}_-^{ik}-\dddot{Q}_{+ik}\dot{Q}_{-}^{ik}}}\nn\\
&&+\frac1{30}I_-Q_{+ik}^{(4)}Q_{+ik}^{(4)}-\frac1{15}\ddot{I}_-\dddot{Q}_{+ik}\dddot{Q}_{+ik}+\dddot{I}_-\pa{\frac8{15}\dddot{Q}_{+ik}\ddot{Q}_{+ik}-\frac25Q_{+ik}^{(4)}\dot{Q}_{+ik}}+\frac1{45}I_-^{(4)}Q_{+ik}^{(4)}Q_{+ik}\nn\\
&&\left.-\frac2{27}\pa{I_- I_+^{(4)}-2I_-^{(4)}I_+}I_+^{(4)}+\frac29\pa{\ddot{I}_- \dddot{I}_+-2\dddot{I}_- \ddot{I}_+}\dddot{I}_+\right\}+{\cal S}_{\rm nl}\,,
\ee
where the non-local term ${\cal S}_{\rm nl}$ of the action is the same reported in \cite{Porto:2024cwd}.
Finally, we report the result for the two bottom contact diagrams\footnote{The traceless part has previously been computed in  \cite{Blumlein:2021txe} and \cite{Almeida:2022jrv}, but without fully taking into account the contributions to the quadratic coupling reported in eq.(\ref{eq:quadratic}), and thus giving incorrect result.}:
\be
\label{eq:actmem_de}
{\cal S}_{\rm cont}&=&G^2\int_t\left\{Q_{+ij}^{(4)}Q_{+jk}\pa{Q_{-}^{ik}}^{(4)}-\frac12Q_{+ij}^{(4)}Q_{+jk}^{(4)}Q_{-}^{ik}+\right.\nn\\
&&+\frac 13I_+Q_{+ik}^{(4)}\pa{Q_-^{ik}}^{(4)}+\frac65\dddot{I}_+\pa{\dddot{Q}_{+ik}\ddot{Q}_-^{ik}-\ddot{Q}_{+ik}\dddot{Q}_{-}^{ik}}\nn\\
&&+I_+^{(4)}\paq{\frac1{45}\pa{Q_{+ik}^{(4)}Q_{-}^{ik}-Q_{+ik}\pa{Q_-^{ik}}^{(4)}}
  +\frac8{15}\pa{\dddot{Q}_{+ik}\dot{Q}_{-}^{ik}-\dot{Q}_{+ik}\dddot{Q}_-^{ik}}}\nn\\
&&-\frac16I_-Q_{+ik}^{(4)}Q_{+ik}^{(4)}+\dddot{I}_-\pa{-\frac23\dddot{Q}_{+ik}\ddot{Q}_{+ik}+\frac8{15}Q_{+ik}^{(4)}\dot{Q}_{+ik}}-\frac1{45}I_-^{(4)}Q_{+ik}^{(4)}Q_{+ik}\nn\\
&&\left.+\frac2{27}\pa{I_- I_+^{(4)}-2I_-^{(4)}I_+}I_+^{(4)}-\frac29\pa{\ddot{I}_- \dddot{I}_+-2\dddot{I}_-\ddot{I}_+}\dddot{I}_+\right\}\,.
\ee
Putting everything together, and reworking some terms by parts, we finally obtain:
\be
\label{eq:SmemSijSQM}
{\cal S}_{Q^3}&=&{\cal S}_{QQ, NLO}+S_{{\rm bulk}}+S_{{\rm cont}}\nn\\
&=&\ds\frac{G^2}5\int_t\left\{\frac12Q_{+ij}^{(4)}Q_{+jk}^{(4)}Q_{-}^{ik}-Q_{+ij}^{(4)}Q_{+jk}\pa{Q_{-}^{ik}}^{(4)}+\frac17\dddot{Q}_{+ij}\dddot{Q}_{+jk}\ddot{Q}_{-}^{ik}+\frac27\dddot{Q}_{+ij}\ddot{Q}_{+jk}\dddot{Q}_{-}^{ik}\right.\nn\\
&&\qquad\qquad\left.-4\paq{\dddot{Q}_{ij}Q_{jk}+\frac13\pa{I\dddot{Q}_{ik}+\ddot{I}\dot{Q}^{ik}}}_+\pa{Q_-^{ik}}^{(5)}\right\}+{\cal S}_{\rm nl}\,.
\ee

As it happens for the waveform, the local part of the above expression is different from the corresponding one reported in eq.~(4.12) of \cite{Porto:2024cwd}, which is reported in the second line above, because we are working within the MPM local frame and multipole definitions.
We recognize in the second line the same term $\delta Q^{ik}_+$  introduced in equation (\ref{eq:comphTT}). This would hint at interpreting the difference between our effective action and the one derived in \cite{Porto:2024cwd}, as due to the same coordinate transformation which connected the waveforms; however the structure in the second line of (\ref{eq:SmemSijSQM}) is not complete and cannot be interpreted in this way. For this reason the extra term gives a non-vanishing contribution to the scattering angle, which thus results different from the value currently accepted in the literature \cite{Dlapa:2022lmu}.

Notice also that the terms containing three moments of inertia canceled in
the final result, a welcome feature as they are not associated with gravitational radiation and we thus expect them to be implicitly included in the near zone dynamics, computed in terms of quasi-instantaneous propagators instead of retarded ones.

\subsection{Energy and angular momentum balance }
We can verify the self-consistency of our findings by computing the energy and angular momentum non-conservation of the binary system, implied by (\ref{eq:SmemSijSQM}).

The result must be equal, modulo total derivatives, known as {\it Schott\ terms}, to the energy and angular momentum fluxes computed from the $h^{TT}_{ij}$ displayed in (\ref{eq:hTT}).

The standard derivation requires the calculation of the equations of motion as
\be
\left.\frac{\delta {\cal S}}{\delta x_-}\right|_{x_-=0}=0\,,
\ee
and the balance equations as
\be
\dot{E}=\mu v^ia_i\,,\quad \dot{L}^i=\mu\epsilon^{ijk}x_j a_k\,.
\ee
For the energy, we find, modulo Schott terms,
\be\label{eq:Edot}
\dot{E}&=&\frac{2G^2}5\bigg[\dddot{Q}^{ij}\dddot{Q}_{ik}\dddot{Q}_{jk}-\frac52\dot{Q}^{ij}Q^{(4)}_{ik}Q^{(4)}_{jk}-\dot{I}Q^{(4)}_{jk}Q^{(4)}_{jk}+2 \dddot{I}\dddot{Q}_{jk}\dddot{Q}_{jk}
 +\frac23I^{(4)}\pa{2\dddot{Q}_{jk}\ddot{Q}_{jk} -Q^{(4)}_{jk}\dot{Q}_{jk}}\bigg]\,,\nn\\
\ee
which is precisely equal to (minus) the energy flux carried at infinity by the gravitational wave (\ref{eq:hTT}) according to the textbook formula
\be
{\cal F}=\frac{r^2}{32\pi G}\int_{\Omega}\langle\dot{h}^{TT}_{ij}\dot{h}^{TT}_{ij}\rangle\,.
\ee
Coming to the angular momentum, the local part of the action (\ref{eq:SmemSijSQM}) gives (still modulo Schott terms)
\be 
\ds \dot{L}^i_{\rm loc}&=&\frac{4G^2}5\varepsilon_{iab}\left\{Q^{(4)}_{ak}\paq{\dddot{Q}_{bl}\dot{Q}_{kl}+\dot{Q}_{bl}\dddot{Q}_{kl}+\frac12Q_{bl}Q^{(4)}_{kl}}+\frac23\paq{Q^{(4)}_{ak}\dddot{Q}_{bk}\dot{I}+Q^{(4)}_{ak}\dot{Q}_{bk}\dddot{I}-\dddot{Q}_{ak}\ddot{Q}_{bk}\dddot{I}}\right\}\,,\nn\\
\ee
to which one has to add the contribution from the nonlocal part of the action ${\cal S}_{\rm nl}$, which we report from \cite{Porto:2024cwd}:
\be
\ds \dot{L}^i_{\rm nl}&=&\frac{4G^2}{35}\varepsilon_{iab}\left\{\dddot{Q}_{ak}\ddot{Q}_{bl}\dddot{Q}_{kl}-Q_{ak}^{(3)}\int_{-\infty}^t Q_{bl}^{(3)}(t')Q_{kl}^{(3)}(t')dt'\right\}\,.
\ee
The sum,
\be\label{eq:Ldottot}
\ds \dot{L}^i&=&\frac{4G^2}5\varepsilon_{iab}\left\{Q^{(4)}_{ak}\paq{\dddot{Q}_{bl}\dot{Q}_{kl}+\dot{Q}_{bl}\dddot{Q}_{kl}+\frac12Q_{bl}Q^{(4)}_{kl}}+\frac17\dddot{Q}_{ak}\ddot{Q}_{bl}\dddot{Q}_{kl}
-\frac17Q_{ak}^{(3)}\int_{-\infty}^t Q_{bl}^{(3)}(t')Q_{kl}^{(3)}(t')dt'\right.\nn\\
&&\qquad\qquad\qquad\left.+\frac23\paq{Q^{(4)}_{ak}\dddot{Q}_{bk}\dot{I}+Q^{(4)}_{ak}\dot{Q}_{bk}\dddot{I}-\dddot{Q}_{ak}\ddot{Q}_{bk}\dddot{I}}\right\}\,,
\ee
coincides with (minus) the angular momentum flux at infinity
\be
{\cal F}_{L^i}=\frac{r^2}{32\pi G}\varepsilon_{ikl}\int_{\Omega}\langle\dot{h}^{TT}_{ab}x^k\partial^lh^{TT}_{ab}+2\dot{h}^{TT}_{ak}h^{TT}_{al}\rangle\,, 
\ee
 thus sealing the self-consistency of our derivation.
 
 The corresponding fluxes usually reported in the MPM literature \cite{Arun:2007sg, Arun:2009mc} have a slightly different form because of the already discussed implicit 2.5PN contribution contained in the MPM quadrupole term, but they are identical to (\ref{eq:Edot},\ref{eq:Ldottot}), modulo Schott terms, when this contribution is taken into account.

\section{Conclusions}
\label{sec:disc}

We have re-derived the quadrupole-quadrupole gravitational waveform within the
Effective Field Theory framework, obtaining the same result as the one
originally obtained within the MPM formalism.
We have understood the origin of the apparent mismatch with the result found in \cite{Porto:2024cwd} as due to the same coordinate transformation that connects the Burke-Thorne to the Damour-Deruelle form of the radiation reaction equation, and we have clarified how this coordinate change is implicitly executed in the EFT treatment.

We also originally obtained a corresponding in-in effective action, whose
dynamics gives balance equations in agreement with the standard expressions for
the energy and angular momentum fluxes associated to gravitational radiation,
thus ensuring the self-consistency of our results. 

The scattering angle derivation in agreement with the (suitably PN-expanded)
post-Minkowskian result of \cite{Dlapa:2022lmu} is however not addressed in
the present work.
As this problem is successfully tackled in \cite{Porto:2024cwd}, we leave
a fully satisfactory treatment of the memory-induced dynamics, providing
{\em both} the MPM waveform {\em and} the 4PM scattering angle, to
future investigations.

\section*{Acknowledgments} 

The authors thank Luc Blanchet and Rafael Porto for useful discussions.
S.F. is supported by the Fonds National Suisse, grant $200020\_191957$, and by the SwissMap National Center for Competence in Research; he also acknowledges the ICTP-SAIFR for kind hospitality during the preparation of this work, with the support of the FAPESP grant 2021/14335-0. A.M. also acknowledges support by FAPESP, under grant 2023/15479-1.
RS wishes to acknowledge FAPESP grants n. 2022/06350-2 and 2021/14335-0,
as well as CNPq grant n.310165/2021-0.
GLA acknowledges support from the National Natural Science Foundation of China under Grant No. 12247103.

\appendix
\section{Ward identities in Magnetic Quadrupole-Angular momentum amplitude}\label{app:magLtail}
\begin{figure}
	\begin{center}
		\begin{tikzpicture}
      			\draw [black, thick] (0,0) -- (4.,0);
      			\draw [black, thick] (0,-0.1) -- (4.,-0.1);
      			\draw [red, densely dotted, line width=1.5] (3,0) -- (2,1.5);
			\draw[decorate, decoration=snake, line width=1.5pt, green] (.5,0) -- (2,1.5);
			\draw[decorate, decoration=snake, line width=1.5pt, green] (2,1.5) -- (3.5,2) ;
      			\filldraw[black] (3.,-0.1) circle (3pt) node[anchor=north] {$J^i$};
      			\filldraw[black] (.5,-0.1) circle (3pt) node[anchor=north] {$J^{kl}$};
		\end{tikzpicture}\hspace{1cm}
	\end{center}
	\caption{L-tail diagram involving a magnetic quadrupole.}
	\label{fig:Ltail}
\end{figure}
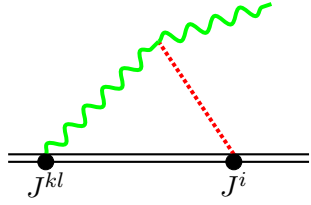

The self-consistency of the EFT framework presented in this paper can also properly justify the result obtained in Ref.~\cite{Almeida:2023yia} for the amplitude of the angular momentum failed-tail involving a magnetic quadrupole-angular momentum interaction, whose diagram is depicted in Fig.~\ref{fig:Ltail}. In that reference, we resorted to methods lying outside the scope of field theory, following a procedure widely utilized in MPM computations, to correct the emission amplitude for such a process. As we discusse there, this amplitude initially failed to be consistent with the Ward identity as its spatial part read:
\be
k^\mu {\cal A}^{(m-L-tail)}_{\mu l}=-\frac4{15}G \omega^5 L^k J^{kl}\pa{\omega}\,.
\ee
Now, in light of the framework presented in this paper, in addition to the original emission amplitude, one should also consistently consider the LO amplitude ${\cal A}^{\rm LO}_{0i}=-\frac12 P^i+\dots$, which then implies
\be
k^\mu\pa{{\cal A}^{LO}+{\cal A}^{(m-L-tail)}}_{\mu l}=0\quad\Rightarrow \quad \dot{P}^i =\frac8{15}G L^k J_{kl}^{(5)}\,.
\ee
In particular, this results agrees with (3.11) of \cite{Blanchet:2018yqa} and it incorporates the right stress-energy conservation law, which, when working at NLO, is given by $\pa{T^{\mu\nu}+\tau^{\mu\nu}}_{,\nu}=0$, and not just $T^{\mu\nu}_{,\nu}=0$.

\section{Schott terms}\label{app:Schott}
Using the radiation reaction acceleration (\ref{eq:aRR}) one obtains
\be
\label{eq:SchottM}
\ba{rcl}
\dot M&=&\ds\mu{\bf v}\cdot {\bf a_{SH}}\\
&=&\ds -\frac 15\dot Q_{ij}Q_{ij}^{(5)}+\frac 12\frac d{dt}\paq{3Q^{(5,0)}+7Q^{(4,1)}+
  5Q^{(3,2)}}\,,\\
&=&\ds -\frac 15\dot Q_{ij}Q_{ij}^{(5)}+\dot M_{Schott}\,,
\ea
\ee
where the last line serves as a definition of the Schott terms in the energy
$M_{Schott}$, whose explicit expression in terms of orbital variable is
\be
\label{eq:SchottM_exp}
M_{Schott}=8\nu^2\frac{G^2M^3}{r^2}\dot r\pa{\dot r^2-v^2}\,.
\ee
This forms agrees with the one obtained by varying the
Newtonian energy via transformation (\ref{eq:deltar}):
\be
\label{eq:M2p5}
M_{Schott}&=& \mu\paq{ {\bf v}.\delta\dot{{\bf r}}-{\bf a}_N.\delta{\bf r}}
\nn\\
\implies \dot M_{Schott}&=&\ds\mu\paq{{\bf a}_{RR}.\delta\dot{{\bf r}}+ {\bf v}.\delta\ddot{{\bf r}}-v^j\partial_j a^i_N \delta r^i}=- \mu{\bf v}.\delta{\bf a}_{RR} + {\cal O}\pa{5PN}\,.
\ee

Analogously for the angular momentum one has
\be
\ba{rcl}
\dot L^{i|j}&=&\ds\mu \pa{x^ia^j_{RR}-x^ja_{RR}^i}\\
&=&\ds-\frac 45Q_{k[i}Q_{j]k}^{(5)}+2\frac d{dt}
\pa{\alpha Q_{k[i}Q^{(4)}_{j]k}+\frac{5\alpha-4}3\dot Q_{k[i}\dddot Q_{j]k}}\,,
\ea
\ee
where $\alpha$ is an arbitrary coefficients that cancel when expressing the quadrupole in terms of orbital variables, as
\be
\ba{rcl}
\ds L^{ij}_{(Schott)}&\equiv&\ds 2\alpha Q_{k[i}Q^{(5)}_{j]k}+
2\frac{5\alpha-4}3\dot Q_{k[i}Q^{(4)}_{j]k}\\
&=&\ds -8\nu^2\frac{G^2M^3}{r^2}\dot r \pa{x^iv^j-x^jv^i}\,.
\ea
\ee
As for the case of the energy, the Schott term for the angular momentum
can be ascribed to the change of variable
\be
\dot{L}^{ij}_{Schott}=2\mu\paq{\delta r^{[i} a^{j]}+r^{[i}\delta\ddot{r}^{j]}}\,,\quad 2\mu r^{[i}\delta a^{j]}_{RR}\simeq 2\mu r^{[i}\paq{\delta r^k \partial_k a_N^{j]}-\delta \ddot{r}^{j]}}\,.
\ee


\end{document}